\documentclass{jfm}
\usepackage[utf8]{inputenc}
\usepackage[T1]{fontenc}
\usepackage{graphicx}
\usepackage{xcolor}
\usepackage{amsmath}
\usepackage{ulem}
\usepackage{breqn}
\usepackage{epstopdf}
\usepackage{gensymb}

\let\vec\mathbf
\let\overrightarrow\mathbf

\newcommand{\ch}{}

\begin{document}

\newtheorem{lemma}{Lemma}
\newtheorem{corollary}{Corollary}
\shorttitle{Energy budget in internal wave attractors experiments} 
\shortauthor{G. Davis et al.} 

\title{Energy budget in internal wave attractor experiment\ch{s}}

\author
 { G\'eraldine Davis\corresp{\email{Geraldine.Davis@ens-lyon.fr}}
 ,
  Thierry Dauxois\corresp{\email{Thierry.Dauxois@ens-lyon.fr}}
 ,
  Timothée Jamin\corresp{\email{Timothee.Jamin@ens-lyon.fr}}
  \and 
 Sylvain Joubaud \corresp{\email{Sylvain.Joubaud@ens-lyon.fr}}
  }

\affiliation
{
\aff{}
Univ Lyon, ENS de Lyon, Univ Claude Bernard, CNRS,
Laboratoire de Physique, F-69342 Lyon, France
}

\maketitle

\begin{abstract}
\ch{The current paper} presents an experimental study of the energy budget of a two-dimensional internal wave attractor in a trapezoidal domain filled with uniformly stratified fluid. The injected energy flux and the dissipation rate are simultaneously measured from a \ch{two-dimensional, two components,} experimental velocity field. The pressure perturbation field needed to quantify the injected energy is determined from the linear inviscid theory. The dissipation rate in the bulk of the domain is directly computed from the measurements, while the energy sink occurring in the boundary layers are estimated using the theoretical expression of the velocity field in the boundary layers, derived recently by Beckebanze \textit{et al.} (J. Fluid Mech. 841, 614 (2018)). In the linear regime, we show that the energy budget is closed, in the steady-state and also in the transient regime, by taking into account the bulk dissipation and, more important, the dissipation in the boundary layers without any adjustable parameters. The dependence of the different sources on the thickness of the experimental set-up is also discussed. In the nonlinear regime, the analysis is extended by estimating the dissipation due to the secondary waves generated by triadic resonant instabilities showing the importance of the energy transfer from large scales to small scales. \ch{The method tested here on internal wave attractors can be generalized straightforwardly to any quasi two-dimensional stratified flow.}
\end{abstract}

\section{\label{sec:intro}Introduction}
In fluid mechanics, \ch{energy budgets provide}  useful insights on the mechanisms 
at play in the energy cascade, the transfer from the large scale of energy injection to the small scales 
where dissipation is taking place. In geophysical and astrophysical systems, this is of paramount importance to understand the role played by 
internal gravity waves, generated through different mechanisms at large scale, in the induced mixing of the fluid occurring at small scales~\citep{Munk1966, MunkWunsch1998,Ivey2008,Mathis2019,SutherlandPRF2019}.

Recently, there \ch{has} been a lot of interest in the determination of the injected energy flux {into the internal wave field} which requires the simultaneous measurement of both  velocity and  pressure fields.
Direct calculations can be rather straightforwardly performed in numerical simulations, since both fields are computed~\citep{Lamb2004,MathurPeacock2009,Rapaka2013}. This is usually much more complicated in laboratory experiments \ch{and even more in} field measurements \ch{where one can have at best the information on a one dimensional line~\citep{HanVanHaren}}. Different methods have been however proposed depending on the experimental techniques used to visualize the internal waves. \cite{Lee2014} require only the velocity field measured using the particle image velocimetry technique (PIV). Density perturbations obtained with the synthetic Schlieren technique can lead to the estimation of the energy flux for linear~\citep{Clark2010, Allshouse2016} or nonlinear stratifications~\citep{Lee2018}. Finally, \cite{Nash2005} present the method to compute the energy flux from ocean observations.

The counterpart of the energy \ch{sources} is the energy sinks. In the ocean, it has been observed that the loss of energy can occur very far from the energy sources. Even if some energy is lost \ch{during propagation}, some dissipation occurs near topographies indicating that \ch{what is} happening near boundaries clearly matters in the global energy budget~\citep{Nikurashin2010}. Indeed, small streamwise undulations of a bottom topography  may have a\ch{n } effect on the structure of the viscous boundary layer~\citep{Passaggia2014}
\ch{and can induce a decay of internal tides in the ocean~\citep{BulherMiranda}}. In laboratory experiments, \ch{the propagation of an internal wave beam is affected by viscous damping}  
as initially studied by~\cite{ThomasStevenson1973} or \ch{ through} instabilities of internal wave beams~\citep{DauxoisARFM2018}. It has been also  recently shown that viscous boundary layers can induce strong mean flow due to streaming~\citep{Horne2019, Renaud2019}.

In a recent paper,~\cite{Beckebanze2018} investigate the role of the boundary layers \ch{on} the resulting spectrum of \ch{an internal wave field attractor}. The latter for stratified fluids or inertial wave attractors for rotating fluids are a very interesting scenario \ch{for energy transfer} since energy is linearly transferred  through smaller scales due to the \ch{focusing} on a sloping topography~\citep{DauxoisYoung1999}. Depending on the geometry of the domain, multiple reflections can lead to a concentration of wave energy on a closed loop, the so-called internal wave attractor~\citep{MaasLam1995, MBSL1997}. In the presence of dissipation, the width of the attractor beam is set by the competition between geometric focusing and viscous broadening~\citep{RGV2001,Ogilvie2005,GSP2008,HBDM2008}. \cite{Beckebanze2018} model the different dissipation processes and, in particular, take into account the part due to viscous boundary layers. They finally describe with a very good agreement the spectrum of linear internal wave attractors obtained in previous laboratory experiments~\citep{HBDM2008, Brouzetetal2016}. Furthermore, high concentration of energy \ch{make attractors prone} to different instabilities~\citep{Maas2005}. Several experiments have recently shown the scenario of instability that can occur in internal wave attractors~\citep{SED2013,BrouzetEPL2016,Brouzetetal2017} which involves in particular the triadic resonance instability~\citep{DauxoisARFM2018}. The effect of these nonlinearities on the dissipation rate for the attractor has been studied numerically by~\cite{JouveOgilvie2014}.

 In the present paper, we consider experimentally the complete energy budget for an internal wave attractor in \ch{a} trapezoidal geometry. 
 Measuring the velocity field only in the bulk (i.e. outside the boundary layers), we carefully show how to quantify the injected power and the energy sink. Both the internal shear layer and the boundary layers are taken into account. We analyse the complete energy budget in two linear scenarios  but also in a nonlinear scenario involving a cascade of triadic resonance instabilities.

The paper is organized as follows. Once the experimental set-up used in the laboratory has been described in~\S~\ref{sec:setup}, we introduce in~\S~\ref{sec:model}  the corresponding energy budget model. The injection energy flux and the different terms involved in the dissipation rate are then presented. In~\S~\ref{sec:results}, we first carefully {measure and analyse} the energy balance in a linear case. In a second stage, when considering nonlinear regimes, we demonstrate that secondary waves emerging in the energy cascade have a non negligible part \ch{of} the total dissipation. The importance of the dissipation in the boundary layers is then discussed using a linear internal wave attractor in a different experimental set-up. We finally conclude in~\S~\ref{sec:conclusion} and draw some perspectives.

\section{\label{sec:setup}Experimental set-up}
As shown in figure~\ref{figureschematic}, a rectangular tank of size $\ch{L \times W \times H =} 2000 \times 170 \times 1000 \ \mathrm{mm^3}$ is 
filled with a fluid linearly stratified with salt and for which density \ch{varies} from $1.06\ \mathrm{kg/L}$ at the bottom to $1.00\ \mathrm{kg/L}$ at the surface. The origin $O$ and the coordinates $x$, $y$ (horizontal) and $z$ (vertical) are defined \ch{in} the figure~\ref{figureschematic}. The buoyancy frequency, defined as $N=\sqrt{-({g}/\rho_r)\textrm{d}\rho_0(z)/\textrm{d}z}$ in which $\rho_0(z)$ is the density stratification at rest and $\rho_r$ is the averaged density of the stratified region, is chosen \ch{at $0.68\pm0.02$}~rad/s. A sloping wall, tilted by an angle $\alpha$ with respect to the vertical, delimits a trapezoidal fluid domain of length~$L=1720$~mm (measured along the bottom) and depth~$H=920$~mm. The system is forced by an internal wave generator (left wall) {of the same height $H$} made of a flexible membrane {\ch{pivoting} around an axis} at mid-height. The top and bottom ends remain vertical and are oscillating at
the frequency $\omega_0$ with an amplitude~$a_0$. 
These two parameters can be tuned appropriately by a simple
electro-mechanic device.
The frequency is chosen constant throughout all experiments described here while different amplitudes are used but kept constant during each of them.
The position of the flexible membrane 
 can thus be given with a very good agreement by $a_0\sin(\omega_0 t)\cos(\pi z/H)$, where~$z$ is the vertical coordinate taken from the bottom of the tank. {Due to the shape of the generator, the vertical component of the induced wavevector is given by $m=2\pi/\lambda_z=2\pi/(2H)$ 
while the horizontal one, $\ell=2\pi\omega_0/(N\lambda_z\sqrt{1-(\omega_0/N)^2} )$, is obtained using the dispersion relation~{$\omega_0= N \ell/\sqrt{\ell^2+m^2}=N\sin\theta$, 
with $\theta$  the angle between the wavevector and the vertical. Note that this angle is also equal to the angle between the rays and the horizontal as displayed in figure~\ref{figureschematic}}\ch{,  where \textit{rays} are energy propagation path}.
 The typical wavelength of the injected wave-field  $2H\sqrt{1-(\omega_0/N)^2}$ is of the order of $1500$~mm.}
 
 {The experimental setup has been designed such that the wave-field is quasi-bidimensional. This means that the transverse velocity field $v$ can be considered null and that the horizontal and vertical components of the velocity field, denoted $u$ and $w$, do not depend on the transverse direction $y$, except in the boundary layers. This property has been experimentally and numerically verified for the linear and weakly nonlinear regime  in}~\cite{Brouzetetal2016}. 
 
 The velocity field {($u$, $w$)} measured in the vertical mid-plane ($y=0$) {is} then monitored as a function of spatial coordinates and time, using the 
standard PIV technique. The flow is indeed seeded with {$10$-$\mu$m silver-coated hollow glass spheres of density $1.4$~kg/m$^{3}$, and illuminated by a 532-nm 2W-continuous laser shaped into a vertical sheet. The cross-correlation algorithm~\citep{FinchamDelerce2000} is performed \ch{by} analyzing windows of typical size $31\times 31$ pixels with $20$\% overlap. The final mesh size is approximately 1~cm.

\begin{figure}
\begin{center}
\includegraphics[width=0.5\linewidth]{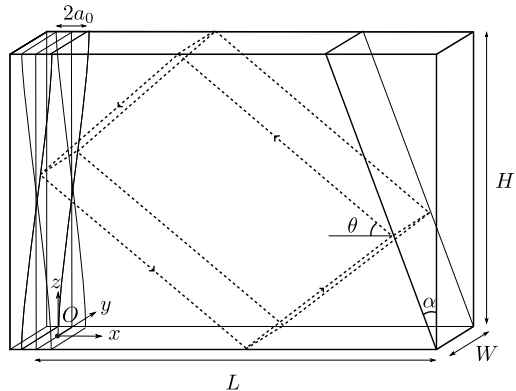} 
\def\svgwidth{0.9\textwidth} 
\caption{Schematic view in perspective of the experimental setup.
A sloping wall, inclined at an angle~$\alpha$ with respect to the vertical, is put 
inside an immobile tank 
delimiting  a trapezoidal fluid domain of length~$L=1720$~mm (measured along the bottom), depth~$H=920$~mm and thickness $W=170$~mm.
A flexible membrane is oscillating around an axis located at mid-height, \ch{producing an horizontal velocity field $a_0\sin(\omega_0t)\cos(\pi z/H)$.}
Black dotted lines show \ch{an internal wave} billiard geometric prediction of an attractor of frequency $\omega_0$.
The angle {$\theta$} of the first branch {(top right)} is given by the dispersion relation $\omega_0=N\sin\theta$. {The origin $O$ is taken at the bottom left part of the {fluid domain}, at mid-width.} }
\label{figureschematic}
\end{center}
\end{figure}

Figure~\ref{figure_typ_PIV} presents a typical experimental measurement of the amplitude of the velocity field measured in the steady state and within the linear regime.
 One clearly sees that the velocity field is focused around the infinitely thin inviscid theoretical attractor, depicted by the dashed line,
 with counterclockwise energy propagation.
Note that the PIV images show a different pattern from the ones obtained using synthetic Schlieren~\citep{SED2013,Brouzetetal2016}. Indeed, this last technique, being sensible to the gradient of the density, emphasizes the small scales. As one is interested in the energy budget, the PIV technique is more appropriate \ch{as we will consider the viscous dissipation, which depends only on the velocity field}.

 The finite width $\lambda$ of the attractor (approximately 100~mm) is the result of the equilibrium between \ch{focusing} after reflection on the slope and the viscous spreading. The amplitude, which is at least five times larger than the wave-maker velocity $a_0 \omega_0$, attains its maximum
 just after the focalization on the inclined slope: this is the start of the first branch of the attractor. The amplitude gradually decreases along the attractor to end
 in the fourth branch, of small amplitude and with a rather large width. 
 \ch{After} reflection on the slope, the fourth branch is focalized again and one recovers
 the first branch, thin and with a large amplitude. The dissipation along the attractor {is} precisely balanced by the focalization and one gets a stable limit cycle,
 the internal wave attractor.

\begin{figure}
\begin{center}
\includegraphics[width=0.5\linewidth]{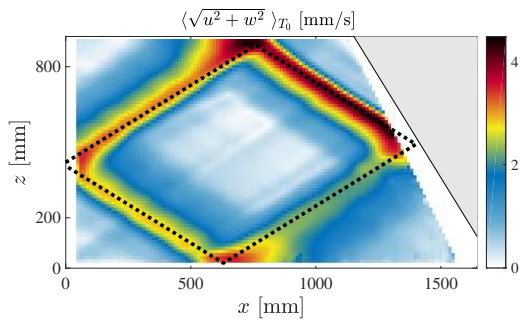} 
\caption{Spatial evolution within the tank of the amplitude of the velocity field $\langle \sqrt{u^2+w^2}\rangle_{T_0}$, averaged over 10 forcing periods $T_0=2\pi/\omega_0$
around $t=50~T_0$.
 The theoretically expected  attractor for this geometry is  depicted  with  the  dashed line.
The parameters of the experiment are $a_0=2$~mm, $\omega_0/N=0.52$, $N=0.68\ch{\pm0.02}$~rad/s and $\alpha=29.5\degree$.
{Note that the shadow of the slope modifies the illumination of the laser sheet and therefore hinders the measurement of the velocity field in the region very close to the slope: this is why a mask (in white in the figure) with an angle different from the slope angle
has been chosen to avoid ambiguous measurements in this region.}}
\label{figure_typ_PIV}
\end{center}
\end{figure}

Now that the experimental set-up has been presented in detail, we will carefully discuss the different contributions involved
in the energy budget of internal wave attractors experiments.

\section{Model\label{sec:model}}
\subsection{Energy budget}
Let us consider an incompressible, nonrotating, stratified fluid in Cartesian coordinates
$(\vec{e}_{x},\vec{e}_{y},\vec{e}_{z})$, where $\vec{e}_{z}$ is the direction opposite to gravity.
In the framework of the Boussinesq approximation,
the equations of motion can be written as 

\begin{align}
\partial_{t}\vec{{u}}+{\vec{u}\cdot\nabla\vec{u}}&=-\frac{1}{\rho_r}\overrightarrow{\nabla}P+b\, \vec{e_{z}}+\nu\nabla^2\vec{u}  \label{linNavierStokes} \\
\partial_{t}b+{\vec{u}\cdot\nabla b}&=-wN^{2} , \label{massconservation}\\
\overrightarrow{\nabla}\cdot\vec{u}&=0
\end{align} 

in which $\vec{u}=(u,v,w)$ is the fluid velocity field, 
$\rho_r$  the average density over the stratified region,
$P$  the pressure variation with respect to the hydrostatic equilibrium pressure,
{$b= g(\rho_0-\rho)/\rho_r$}~the perturbed buoyancy field, 
$N$  the buoyancy frequency
 and $\nu$  the kinematic viscosity.

In \eqref{massconservation}, we have neglected the molecular diffusivity, which would imply a term $ D\nabla^2b$   with $D$ 
 the diffusion coefficient of the stratifying element. 
As we discussed, laboratory experiments use salt as stratification agent, so $D\simeq 10^{-9}$ m$^2$/s which corresponds to the Schmidt number $\nu/D\sim10^{3}$. Therefore {this term} can be neglected with respect to the viscous term appearing in \eqref{linNavierStokes}. 

\ch{Let us recall how it is possible to establish the energy budget of this system as a balance between the injected power and the dissipation rate~(see for example \cite{kundu}).}
Multiplying \eqref{linNavierStokes} with $\vec{u}$ and integrating over ${\cal V}$, the whole volume of the tank (bounded by the surface ${\cal S}$) and using \eqref{massconservation}, one gets

\begin{dmath}
\partial_{t}\int_{\cal V}\mathrm{dV}\left(\frac{\vec{u}^{2}}{2}+\frac{1}{N^{2}}\frac{b^{2}}{2}\right) =-\frac{1}{\rho_{r}}\int_{\cal V}\mathrm{dV}\ \vec{u}\cdot\overrightarrow{\nabla}P - {\int_{\cal V}\mathrm{dV}\ \vec{u}\cdot\overrightarrow{\nabla}\left(\frac{\vec{u}^{2}}{2}+\frac{1}{N^{2}}\frac{b^{2}}{2}\right)} +\nu\int_{\cal V}\mathrm{dV}\ \vec{u}\cdot\nabla^2\vec{u}\,.\label{equationpbypartsvolu}
\end{dmath}
\ch{where $\int_{\cal V}$ and $\partial_t$ have been commuted.} \ch{Indeed, the length of the tank varies only by $\ch{\Delta} L/L~\sim 0.1\%$, with a typical time of variation $t_L=T_0$. 
We will see \ch{in figure~\ref{figure_tf_mes2} in \S~\ref{sec:results}} that the integrand increases from zero \ch{to its final value} on a typical time $t_e\sim 10\ T_0$,
such \ch{that} $\ch{\Delta} L/(L t_L) \ll 1/t_e$. }
\ch{The variations of the frontiers of ${\cal V}$ are thus sufficiently small and slow to be neglected.}

The left-hand-side is the variation of the total energy  $ E_{\textrm{tot}} $ 
that corresponds to the sum of the kinetic and potential energies within the fluid. Note that the latter is defined with respect to the reference stratification which is supposed fixed, \ch{no mixing {being} taken into account}.

Let us show that the right-hand-side corresponds to the sum
of the energy injection rate and the viscous dissipation rate.
We will first focus on the monochromatic case, before considering the polychromatic one in \S~\ref{sec:weakly_non_linear}.
First of all, we turn the volume
integral into a surface integral, keeping in mind that in an
incompressible flow the velocity field satisfies the continuity
equation $
\overrightarrow{\nabla}\cdot\vec{u}=0 .$ Denoting $\vec{\mathrm{d}S}$  the element of surface \ch{which normal} is oriented \ch{out of} the fluid, the first term can therefore
be rewritten as 
\begin{eqnarray}
-\int_{\cal V}\mathrm{dV}\ \vec{u}\cdot\overrightarrow{\nabla}P
&=&\int_{\cal V}\mathrm{dV}\ P\overrightarrow{\nabla}\cdot\vec{u}\ -\oint_{\cal S} P\vec{u}\cdot\vec{\mathrm{d}S} 
\\
&=& -\int_{{\cal S}_g} P\vec{u}\cdot\vec{\mathrm{d}S},
\end{eqnarray}
using the property that the component of the velocity normal to the surface ${\cal S}$ vanishes everywhere except on  ${{\cal S}_g}$, the surface of the {wave generator}
where the  {flexible membrane} injects energy 
in the tank. 

In the same way, the second term of the right-hand-side of \eqref{equationpbypartsvolu} is equal to the energy flux at the boundaries of the domain. Using similar arguments, only the surface of the generator has to be considered. 
 Both quantities {$\vec{u}^{2}\vec{u}$ and $b^{2}\vec{u}$} \ch{thus vanish when averaged over} the forcing period $T_0=2\pi/\omega_0$, as they are odd power of a trigonometric function.
The second term of the right-hand-side of \eqref{equationpbypartsvolu} is thus equal to zero.

Turning again the volume integral into a surface integral, 
 the last term on the right-hand-side of \eqref{equationpbypartsvolu} yields

\begin{dmath}
\ch{\nu} \int_{\cal V}\mathrm{dV}\ \vec{u}\cdot\nabla^2\vec{u}
=-\ch{\nu} \int_{\cal V}\mathrm{dV}\ \left\vert 
 \overrightarrow{\nabla}\vec{u} \right\vert^2\ 
  +\ch{\nu} \oint_{\cal S} u_i \partial_j u_i \mathrm{d}S_j.\label{equationavecleftvert}
\end{dmath}
The last term vanishes when considering fixed solid boundaries {(right slope and bottom)} using the no-slip and no-penetration conditions. It vanishes also at the surface, due to the free-slip condition and because there is no normal variation of the normal velocity{, even in the presence of small surface waves}. For the remaining boundary \-- the generator~\--,  the velocity field corresponds essentially to a  mode-1 induced by the generator itself. Both quantities $u\partial_x u$ and $w\partial_x w$ have thus a vanishing average on the forcing period $T_0=2\pi/\omega_0$ {while} $v=0$.

Equation~(3.3) is thus {re}written as the energy budget within the tank

\begin{align}
\partial_t e_{\textrm{tot}} =  p_{\textrm{inj}}-p_{\textrm{diss}},
\label{eq_bilan}
\end{align}
which relates{, per unit} {mass,} the time variation of the total energy $e_{\textrm{tot}}\equiv E_{\textrm{tot}}/{\cal V}$ with the injected power

 \begin{equation}p_{\textrm{inj}} \equiv {-}\frac{1}{\rho_{r}{\cal V}}\int_{{\cal S}_g}P\vec{u}.\mathrm{d}\vec{S}\label{injectedpower}\end{equation} 
 and the \ch{dissipation} rate 

 \begin{equation}p_{\textrm{diss}}\equiv\frac{\nu}{{\cal V}} \int_{{\cal V}}\mathrm{dV} \left\vert \nabla\vec{u}\right\vert^2.\end{equation}
  For convenience, it is useful to introduce the local density of dissipation rate $\epsilon\equiv \nu\left\vert \nabla\vec{u}\right\vert^2$. \ch{A} temporal average over an integer number of \ch{forcing periods} $T_0$ will be considered. The method to determine the time-averaged values of the total energy, the injected power and the dissipated rate {requires} only the velocity field data and will be presented in the following section.

\subsection{Estimation of the different terms from experimental measurements}
\subsubsection{Total energy}
\ch{Unlike rotating inertial waves where no potential energy can be defined,
equipartition of energy between kinetic and potential subparts can be defined for internal waves.
The time variation of the total energy can thus be safely obtained from PIV  that gives accurate measurements of the velocity components~: $\langle e_{\textrm{tot}} \rangle_{T_0}=\int_{\cal V}\mathrm{dV}\, \langle {\vec{u}^{2}}\rangle_{T_0}/{\cal V}$.}

\ch{The velocity field has been shown to be bidimensional \ch{(\textit{i.e.} only two components invariant in the $y$-direction)} except in the boundary layers, of typical width $d = \sqrt{\nu/\omega_0}\sim 1$~mm \citep{Brouzetetal2016}.}
So their total volume, $2 d {\cal A} + d {\cal L} W$ (where ${\cal L}$  is the perimeter of the \ch{vertical trapezoidal section of the domain}, {$\cal A$ its area} and $W$ the width of the experimental tank) \ch{represents} only $1$\% of the total volume of the tank. The measured
value of the total energy will thus be computed as $\langle e_{\textrm{tot}} \rangle_{T_0}=\int_{{\cal A}
}\mathrm{d}x\mathrm{d}z\, \langle {\vec{u}^{2}}\rangle_{T_0}/{{\cal A} 
}$, \ch{where $\vec{u}$ is measured in the mid vertical plane only.}
 
Because of some experimental issues {near boundaries}, the measured area  is a bit smaller than the total area of the trapezoidal domain. However since this difference is small we assume that the sampled area is representative enough so that our measure of the average energy density is correct.

\subsubsection{Injected power\label{sec:injected_power}}

As the generator is everywhere almost vertical, one can safely approximate \eqref{injectedpower} 
by

\begin{align}
 p_{\textrm{inj}} \simeq \frac{1}{\rho_{r}{\cal A}}\int_{z}Pu\  \mathrm{dz}.\label{injectedpowerapprox}
\end{align}
The determination of the injected power requires therefore the simultaneous measurement of the horizontal velocity field and the pressure near the wavemaker on the left boundary of the tank ($x=0$).
\ch{This kind of generator forces a purely propagating mode-1 with} an amplitude $u_0$ slightly lower than $a_0\omega_0$ depending on the efficiency of the wave maker (see~\cite{MMMGPD2010}),
\begin{eqnarray}
u^{\mathrm{M}1}(x,z,t)&=&u_{0}\cos\left(mz\right)\, \cos\left(\omega_0 t-\ell x\right)\label{mode1composanteu}.
\end{eqnarray}
Using the incompressibility equation and assuming that the flow is two-dimensional, one can straightforwardly get the vertical component,
\begin{eqnarray}
w^{\mathrm{M}1}(x,z,t)&=&{-}u_{0}\frac{\ell}{m}{\sin}\left(mz\right)\, \sin\left(\omega_0 t-\ell x\right),\label{mode1composantew}%
\end{eqnarray}
induced in the fluid by the generator. The associated pressure field can be  deduced by projection of the linear inviscid version of \eqref{linNavierStokes} on the $x$-axis, followed by a straightforward integration

\begin{align}
P^{\mathrm{M}1}(x,z,t)&=\rho_{r}u_{0}\frac{\omega_0}{\ell}{\cos}\left(mz\right)\cos\left(\omega_0 t-\ell x\right), 
\label{puremodeonepressure}
\end{align}
in which the constant of integration 
is set to zero since $P$ accounts only for  the pressure variation with respect to the hydrostatic equilibrium pressure.
The time-averaged theoretically injected power can then be computed as

\begin{equation}
\langle p_{\mathrm{inj}}^{\mathrm{M1}}\rangle_{T_0}
=u_{0}^{2}N\frac{H^{2}}{4\pi}\sqrt{1-\left(\frac{\omega_0}{N}\right)^2}\ \frac{1}{{\cal A}}. \label{calculinjectedpowerpuremode1}
\end{equation}
{This theoretical development is valid for a pure \ch{propagating} mode-1. However, such approximation may be questioned in the presence of the slope, since the size of the closed domain has been chosen to get strong reflected beams, which is required to the formation of the attractor. Some feedback on the forced wave-field from its reflection on the slope is then expected.

To go beyond, it is possible to use the PIV measurements to compute both the pressure and horizontal velocity appearing in~\eqref{injectedpowerapprox}. Indeed, if the PIV technique gives directly the horizontal velocity field, the pressure field can be obtained from~$u$ and~$w$ as follows. 
Since the width of the attractor~$\lambda$, \ch{which is the typical length of variation of the velocity field in the transverse direction of the branch,} is much larger than $d$, we use the {linear and} inviscid
  approximation of \eqref{linNavierStokes} {and \eqref{massconservation}} to get the experimental pressure gradient, which is then 
   integrated using the  {Matlab$"^\copyright$ code} \textit{intgrad2}, available on {\tt www.mathworks.com/matlabcentral/fileexchange/}. The unknown constant of integration necessary for this last step
 is not important since mass conservation implies $\int_{{\cal S}_g}u\ \mathrm{d}y\mathrm{d}z=0$, leading to a vanishing contribution to the injected power. 
 Thus we arbitrarily choose $\langle P \rangle_{\cal A}=0$.

An example of the computed pressure field is presented in figure~\ref{figure_pression}(a).    The pressure field is different from the pure mode-1 pattern~\eqref{puremodeonepressure}:
this will have some consequences on the estimation of the injected power. Because of
 the integration, the pressure field emphasizes the large scale motion contrary to the density gradient or velocity fields
that show the attractor {(see for example figure~\ref{figure_typ_PIV})}. 
\begin{figure}
\begin{center}
(a)\includegraphics[width=0.45\linewidth]{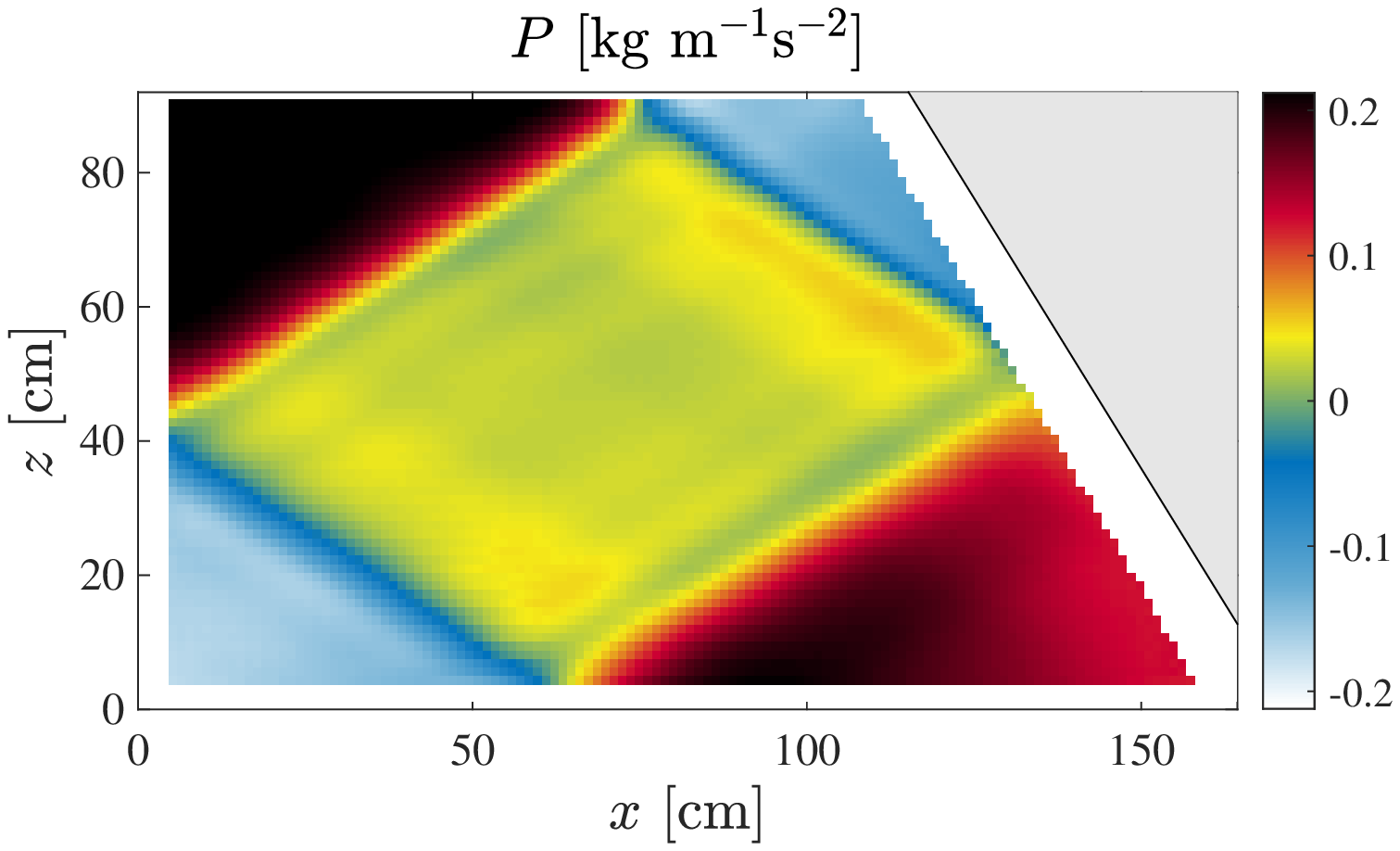} 
(b)\includegraphics[width=0.45\linewidth]{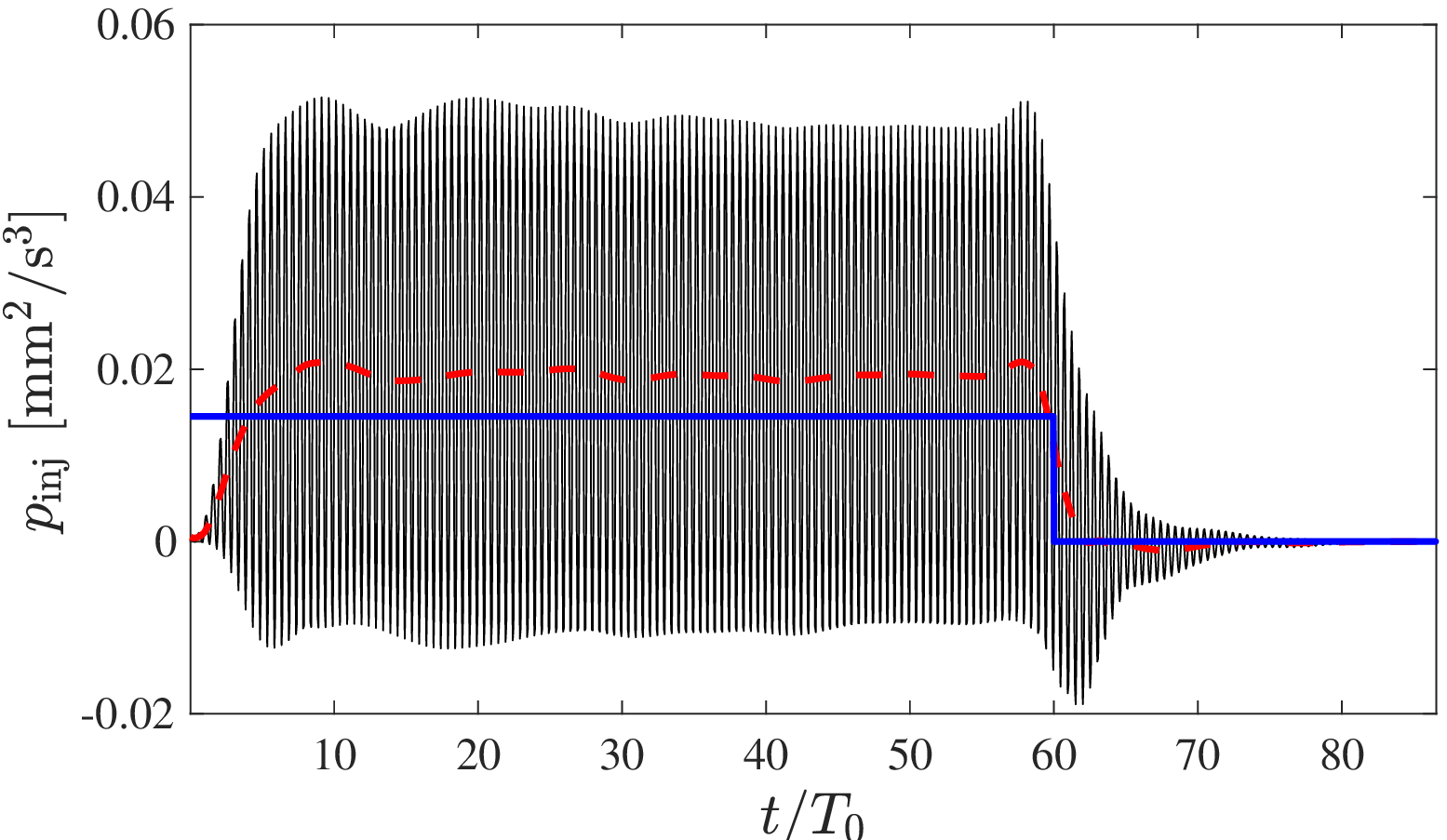} 
\caption{(a) Pressure field $P$, computed via an integration of the experimental velocity fields  $u$ and $w$. 
The {unimportant} constant of integration is arbitrarily chosen to have 
a zero \ch{average pressure field}.
 The pattern is computed {about} $50~T_0$ {after the generator was started}. (b) Evolution in time of the injected power computed at $x_0=5$ cm. 
The oscillating black curve corresponds to the  instantaneous measurement, while 
the red dashed curve is the associated local mean. 
{The generator is switched on at $t=0$ while turned off at $t=60~T_0$.}  
{The thick blue line corresponds to the theoretical injected power~\eqref{calculinjectedpowerpuremode1} expected for a pure mode 1, i.e. neglecting the reflection on the slope.}
{Experimental parameters are those given in the caption of figure~\ref{figure_typ_PIV}.}}
\label{figure_pression}
\end{center}
\end{figure}

Figure~\ref{figure_pression}(b) shows the time evolution of the injected power {$p_{\mathrm{inj}}$} {(solid black line)}  for a typical experiment     {in the linear regime}. 
For visualisation purposes,  this flux of energy  is measured at $x_0\simeq 5$~cm, as  will be discussed  in \S~\ref{Linearregime}.
   {$p_{\mathrm{inj}}$ is oscillating with a mean value {(dashed line)} that first increases until reaching its stationary value. As soon as the generator is turned off, it decreases towards zero. 
  {One notes that the measured value is above {$\langle p_{\mathrm{inj}}^{{\mathrm{M1}}}\rangle_{T_0}$} {(thick blue line), {i.e.}} the theoretical injected power~\eqref{calculinjectedpowerpuremode1} expected for a pure mode 1 neglecting {the finite size of the domain due to the reflection on the slope.}

\subsubsection{Disentangling the different contributions for the dissipative terms}
As already mentioned, the two-dimensional PIV measurements give access to the velocity field in the middle plane $y=0$ and can be extended to the volume except in the boundary layers. 
In the remainder of the paper, this is \ch{named the} bulk part. Since the thickness of the boundary layers  is very small, the velocity gradient may be strong in these regions and the associated dissipative term cannot be neglected. It is therefore convenient to distinguish several different contributions as follows 
\begin{eqnarray}
{p_{\textrm{diss}}}&=&\frac{1}{\cal V}\int_{\cal V}\mathrm{dV} \, {\nu} \left\vert \overrightarrow{\nabla}\vec{u} \right\vert^{2} \\
&=& 
\underbrace{\frac{1}{\cal V} \int_{{\cal V}_{bulk}} \!\!\! \!\!\!\mathrm{dV}\, 
\underbrace{{\nu}\left\vert  \overrightarrow{\nabla}\vec{u} \right\vert^{2}}_{{\epsilon_{\textrm{bulk}}}}}_{{p_{\textrm{bulk}}}}
+ 
\underbrace{
\frac{1}{\cal V} \int_{{\cal V}_{BL,\perp}} \!\!\! \!\!\! \mathrm{dV} \, 
\underbrace{{\nu}\left\vert \overrightarrow{\nabla}\vec{u} \right\vert^2}_{{\epsilon_{\perp}}}
}_{{p_{\perp}}}
+ 
\underbrace{ 
\frac{1}{\cal V} \int_{{\cal V}_{BL,\parallel}} \!\!\! \!\!\!\mathrm{dV} \, 
\underbrace{{\nu}\left\vert\overrightarrow{\nabla}\vec{u} \right\vert^2}_{{\epsilon_{\parallel}}}
}_{{p_{\parallel}}}
.
\end{eqnarray}
\ch{This definition disentangles the bulk dissipation}
 {(denoted with the index \it{bulk}} and that one can expect to measure accurately),
from the contributions of the boundary layers close to the walls.
Among the last ones,
we distinguish 
the dissipation in the boundary layers
along the longitudinal walls (parallel to the \ch{measurement plane} and therefore  perpendicular to  $y${, denoted with the index $\parallel$)},
from   
 the dissipation in the boundary layers perpendicular to the attractor plane {(denoted with the index $\perp$)}.  Let us discuss these three different terms successively.

In the bulk, the transverse velocity $v$ is equal to zero and the velocity components $u$ and $w$ do not depend on $y$. The contribution of the bulk can thus be computed rather straightforwardly from the measurements of the velocity field shown
in figure~\ref{figure_typ_PIV} using the formula

\begin{align}
\epsilon_{\textrm{bulk}} = \nu  \left\vert 
\overrightarrow{\nabla}\vec{u} \right\vert
^{2} = {{\nu}}(\partial_x u)^2 +{{\nu}}(\partial_z u)^2+{{\nu}}(\partial_x w)^2+{{\nu}}(\partial_z w)^2.
\label{eq_epsilon_bulk}
\end{align}

Figure~\ref{epsilon_bulk}(a)  presents the spatial dependence of this quantity in the mid-plane averaged over {10}  $T_0$. As expected, its amplitude is particularly important along the first branch of the attractor, that is located just after the reflection on the slope where the \ch{focusing} {takes} place, and then decreases along the attractor. {The mean of} this quantity on the whole domain leads to~$p_{\textrm{bulk}}$, {the {energy dissipation} rate due to the bulk}.

It is much more difficult to directly estimate the dissipation in the boundary layers. Indeed, since their thickness are very small compared to the size of the tank (around $1$\%),  
it is very hard to 
\ch{measure} the velocity in the  
boundary layers as accurate as in the bulk. 
 A method to evaluate the velocity using the velocity measured in the mid-plane ($y=0$) will then be presented for the two types of boundary layers, parallel to the lateral walls (\S~\ref{sec:diss_parallel}) and perpendicular to {the attractor plane} (\S~\ref{sec:diss_perp}). 
 In both cases, theoretical developments recently reported by~\cite{Beckebanze2018} will be used {and adapted}. For the sake of brevity and clarity, only the main ingredients of their results will be recalled. Notice that linear equations will be used in the two following sections as the Reynolds number is small in viscous boundary layers. 
 \ch{\cite{Brouzetetal2016} showed that the experimental velocity field in the parallel boundary layers is compatible with the theoretical description.}

\begin{figure}
\begin{center}
(a)\includegraphics[width=0.45\linewidth]{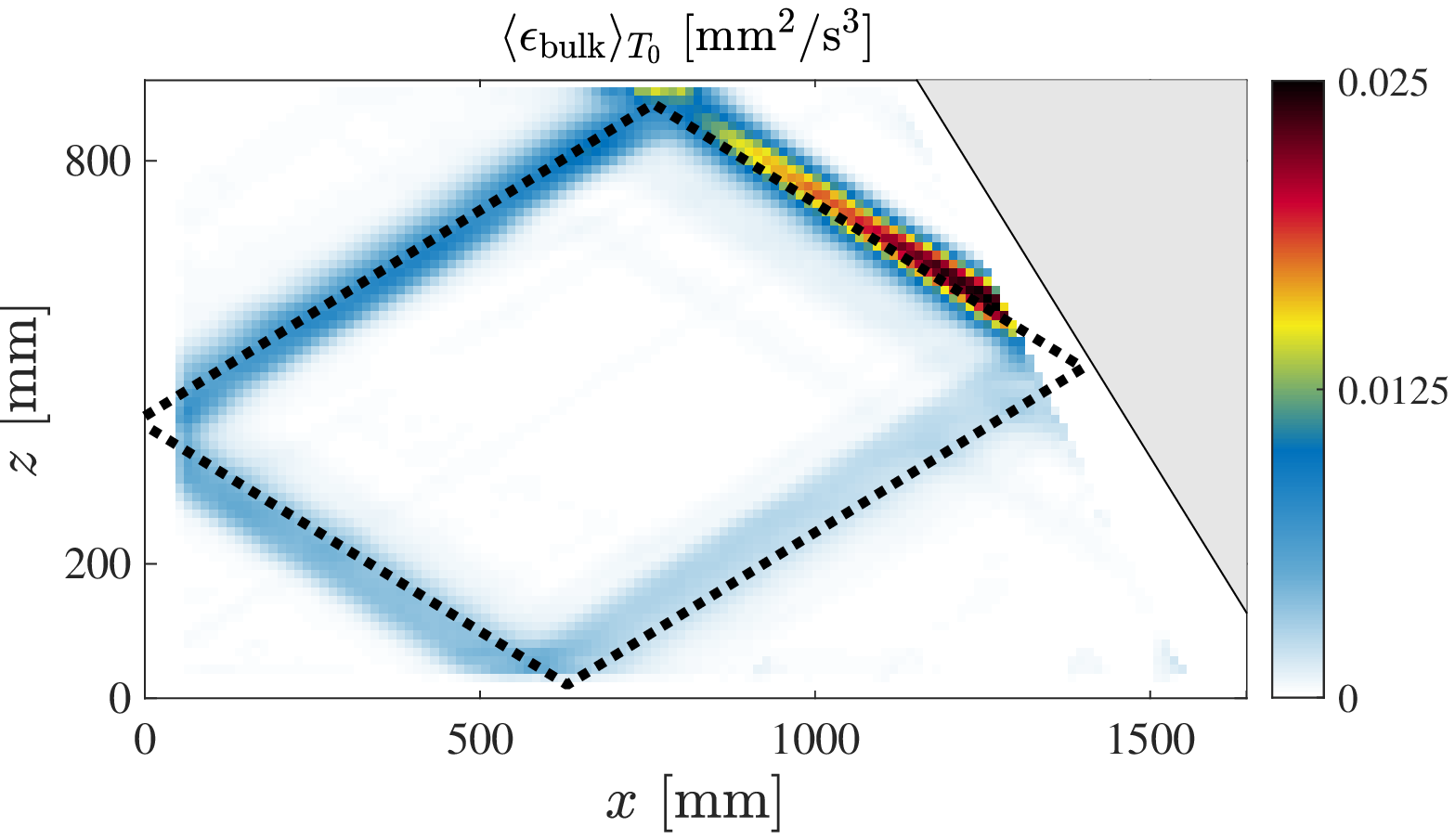} 
(b)\includegraphics[width=0.45\linewidth]{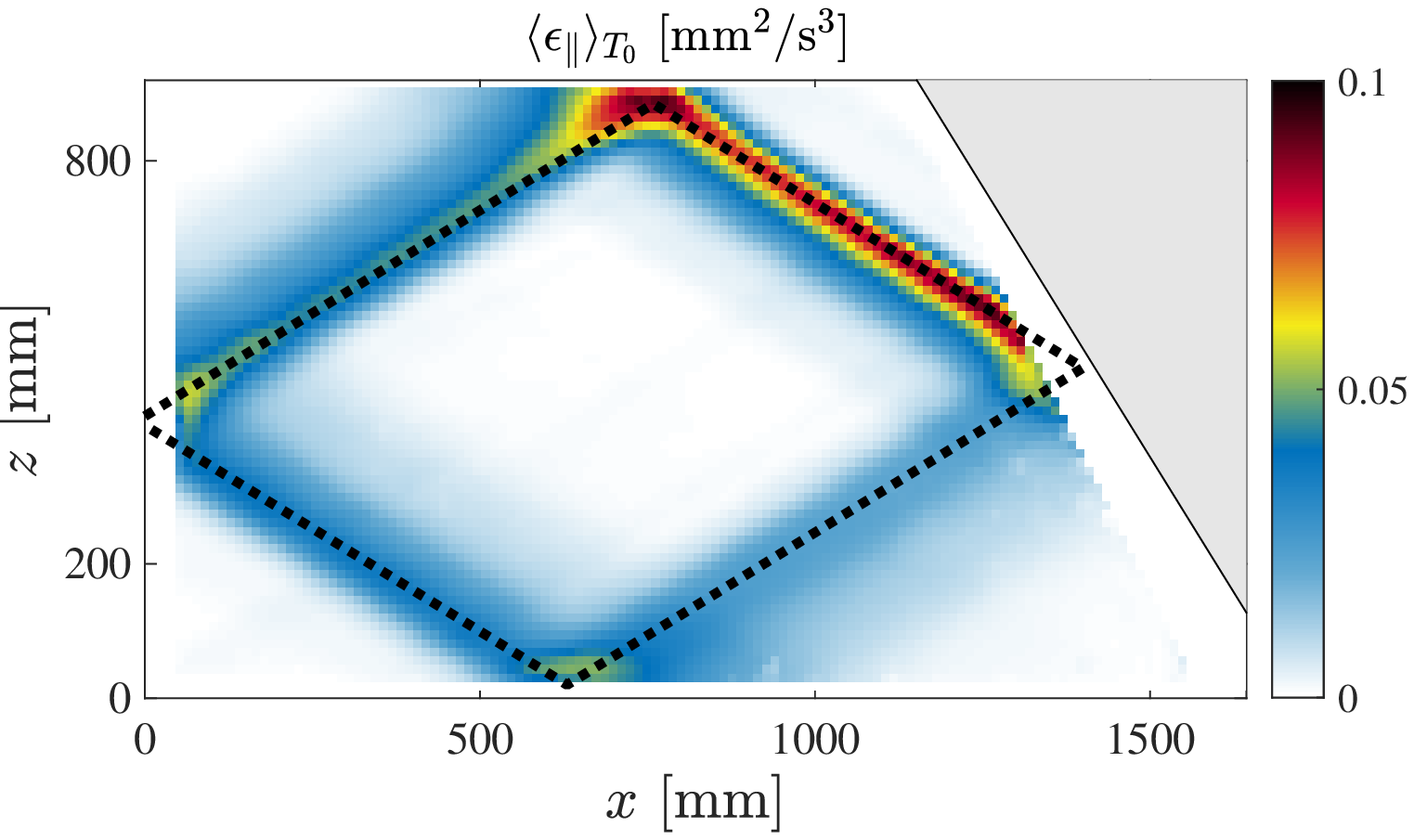} 
\caption{(a) Dissipation rate field within the bulk of the fluid, $\langle \epsilon_{\textrm{bulk}}(x,z)\rangle_{T_{0}}$, measured in the mid-plane. (b) Dissipation rate field in the longitudinal boundary layers, $\langle\epsilon_{\parallel}\rangle_{T_0}$, averaged along the transverse direction. In both panels, the field has been averaged over 10 forcing periods $T_0$ around $t=50~T_0$ an the theoretically expected  attractor for this geometry is  depicted  with  the  dashed line. Experimental parameters are those given in the caption of figure~\ref{figure_typ_PIV}. }
\label{epsilon_bulk}
\end{center}
\end{figure}

\subsubsection{The dissipation in the longitudinal boundary layers $\epsilon_\parallel$\label{sec:diss_parallel}}

Close to both longitudinal vertical walls, the largest contribution to the viscous term~$\nu \vec{\nabla^2} \vec{u}$ comes from the gradients in the $y$-direction. 
Indeed, experimental measurements  show that variations in the $xz$-plane are typically on the centimeter length scale, that is significantly larger 
than the millimeter length scale that appears in the viscous boundary layer.
The evolution of the velocity~\eqref{linNavierStokes} 
and buoyancy fields~\eqref{massconservation}
can {thus} be simplified as 

\begin{align}
\partial_t u &= -\frac{\partial_x P}{\rho_r} + \nu \partial_{yy}u \label{dtwbis},\\
\partial_t w &= -\frac{\partial_z P}{\rho_r}  + b + \nu \partial_{yy}w  \label{dtw} ,\\
\partial_t b &= -  w N^2. \label{dtb}
\end{align}
The transverse velocity $v$ does not vanish any more but its contribution to the dissipation rate remains very small~\citep{Beckebanze2018}. Let us consider a monochromatic internal wave $a = \mathrm{Re} \left[ \underline{a}\ e^{-i\omega_0 t} \right]$ at the frequency $\omega_0$ in which $a$ stands for $u$, $w$, $b$ or~$P$. The case of polychromatic signals will be discussed in \S~\ref{sec:weakly_non_linear}. Introducing the angle~$\theta$ defined through the dispersion relation 
$\omega_0
^2 =N^2\sin^2 \theta$,
\eqref{dtb} can be rewritten as 
$\underline{b}=-i \omega_0  \underline{w}/\sin^{2}\theta$. 
 Equations \eqref{dtwbis} and~\eqref{dtw} can then be  simplified as 

\begin{align}
-i\omega_0\underline{u} &= -\partial_x \underline{P} /\rho_r + \nu\partial_{yy} \underline{u} ,\\
i\omega_0\  \cot^2\theta\ \underline{w} &=-\partial_z \underline{P} /\rho_r+  \nu \partial_{yy}\underline{w},
\end{align}
that have to be solved with the no-slip boundary conditions $ \underline{u}(y={\pm}W/2) =\underline{w}(y={\pm}W/2) =0$, while
$ \underline{u} (y=0)$  and $\underline{w} (y=0)$ can be measured at the center of the tank with the PIV measurements described above.
One finally gets

\begin{align}
\underline{u}(y) &= \left(1-\frac{\cosh\left[i^{1/2}{y}/{d}\right]}{\cosh\left[i^{1/2}{W/(2d)}\right]}\right)\underline{u}(y=0)\label{eq:theory_velocity_long1},\\
\underline{w}(y) &=\left(1-\frac{\cosh\left[i^{-1/2}\cot\theta\, {y}/{d}\right]}{\cosh\left[i^{-1/2} \cot\theta\, {W/(2d)}\right]}\right)\underline{w}(y=0),
\label{eq:theory_velocity_long2}
\end{align}
in which we use $d=\sqrt{{\nu}/{\omega_0}}$, the thickness of the boundary layers.

The dissipation within the longitudinal boundary layers is thus obtained by performing successively the following operations: a time Fourier transform of the wave field in the mid-plane $\vec{u}(x,y=0,z,t)$, a 
band-pass filtering in the Fourier space around the forcing \ch{frequency} $\omega_0$, multiplication by the theoretically derived expressions~\eqref{eq:theory_velocity_long1} and~\eqref{eq:theory_velocity_long2} with respect to the $y$-coordinate to get $\partial_y \underline{\vec{u}}(x,y,z)$.  The real part of the inverse Fourier transform is then taken to generate 
 the signal $\partial_y \vec{u}(x,y,z)$. The value of  $\epsilon_{\parallel}=\nu [ \left(\partial_{y} u \right)^2 +  \left( \partial_{y} w \right) ^2]$ at any point of the domain is then computed and the integration over the whole domain 
 leads to the value of the {energy} dissipation {rate $p_{\parallel}=\int\epsilon_{\parallel} \mathrm{d}V  / {\cal V}$  due to the parallel boundary layers. 
The  $y$-averaged {value} $\langle \epsilon_{\parallel}\rangle_{y,T_0}=(1/W)\int_{-W/2}^{+W/2}\textrm{d}y\,\langle\epsilon_{\parallel}\rangle_{T_0}$ is displayed in figure~\ref{epsilon_bulk}(b) 
  for all points {of the domain}. Here again, the dissipation is larger along the first branch of the attractor and then decreases along the skeleton of the attractor. 
{By comparing figures~\ref{epsilon_bulk}(a) and (b)
, one realizes that t}his boundary layer dissipation is {much} larger than the bulk dissipation. {This difference is by a factor of $10$ when averaged over the entire domain.}  Indeed the $y$-averaged dissipation of the bulk is of the order of 
 $\nu \ (U/\lambda)^2$,
  while the $y$-averaged of the longitudinal boundary layer \ch{dissipation} is of the order 
  $\nu\ (2 d/W)\ (U/d)^2$. One can then estimate the ratio of the two types of  dissipation
${\langle \epsilon_{\textrm{bulk}}\rangle_{y,T_0}/\langle \epsilon_{\parallel}\rangle_{y,T_0}}
\sim {W{d}}/{\lambda^2}\sim 10\%\,$,
{a value} {in} {good agreement with the measured ratio}.

\subsubsection{The dissipation in the transversal boundary layers $\epsilon_\perp$\label{sec:diss_perp}}
The second dissipation term in the boundary layers, $\epsilon_{\perp} $, corresponds to the boundary layers perpendicular to {the attractor plane}, and, more precisely, where a reflection of the internal wave beam occurs. For the present experimental setup, three walls have to be considered while, due to the free slip boundary condition, the reflexion at the top surface is not a source of dissipation. \ch{Note that no linear coupling between internal and surface waves can occur due to the low frequency which leads to a surface wavelength much larger than the size of the set-up.} We have therefore
$p_\perp = p^{[1]}_\perp +  p^{[3]}_\perp + p^{[4]}_\perp $
where $j=1$ stands for the { boundary layer {along} the slope, $j=3$ {close to} the generator {and $j=4$  near the bottom of the tank.}
We will develop the method to estimate the dissipation term associated with a reflection {on a} wall inclined at an arbitrarily angle {$\phi$}, that will stand for $\alpha$, \ch{$\pi$} {or \ch{-$\pi/2$} respectively for $j=1,3,4$.}

{Considering the reflection on a boundary in the inviscid case, the velocity field corresponds to the sum of the incident beam $\vec{u}^{[I]}$ and the inviscid reflected beam $\vec{u}^{[R]}$. It has to fullfill the no{-}penetration condition and is therefore parallel to the {boundary} {along the wall} (free-slip condition). As depicted in figure~\ref{rebond_schema}, we define in the real case the viscous contribution $\vec{u}_\nu=u_\nu \vec{e}_{x}+w_\nu \vec{e}_{z}$ ensuring that the velocity field vanishes along the wall (no-slip condition), yielding the complete velocity field $\vec{u}=\vec{u}^{[I]}+\vec{u}^{[R]}+\vec{u}_\nu$.}
Thus $\vec{u}_\nu$, that has to compensate the {inviscid} free-slip velocity field, is mainly on the $ \vec{e}_{x'}$ direction and 
 may therefore be written as $\vec{u}_\nu= u_\nu' \vec{e}_{x'}$ in the tilted coordinate system $(x',z')$ shown in  figure~\ref{rebond_schema}, with  {$\lim_{z'\rightarrow-\infty}{u}'_\nu=0$. As shown in~\cite{Beckebanze2018}, this is valid at \ch{leading} order in $d/\lambda$ {\--- equal to $1\%$ in our experiment}.

\begin{figure}
\begin{center}
\includegraphics[scale=0.7]{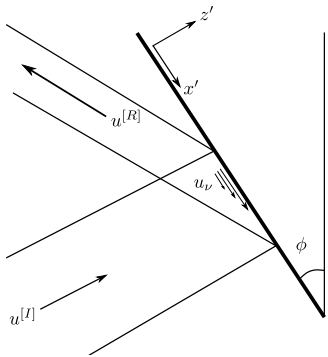} 
\caption{Schematic of a reflection: the incident beam with velocity $\vec{u}^{[I]}$ get bounced on a 
boundary inclined by an anlge $\phi$ with respect to the vertical, leading to the reflected beam with velocity  $\vec{u}^{[R]}$.
The no slip condition induces a viscous velocity field $\vec{u}_\nu$ that counter balances at the boundary the inviscid velocity field $\vec{u}^{[I]} + \vec{u}^{[R]}$. Its exponential decrease as a function of the distance {from} the boundary is schematically represented. Note the introduction of the wall-oriented coordinate system in which $x'$ is the distance along the wall and $z'$ is the distance normal to the wall.}
\label{rebond_schema}
\end{center}
\end{figure}

{In} a boundary {layer}, the dominant part of the viscous term is given by the direction perpendicular to the wall $(z')$, and one has 
${\nabla^2} \vec{u}\approx{\nabla^2} \vec{u}_\nu\approx \nu\partial_{z'z'}\vec{u}_\nu$, as $d\ll\lambda$. {The associated dissipation 
 is thus $\langle p^{[j]}_\perp\rangle_{T_0} =\left\langle    \int_j \mathrm{d}x'\mathrm{d}z'  \left[\partial_{z'}u'_\nu(x') \right] ^2  \right\rangle_{T_0}/{\cal A} $}.  The viscous velocity, pressure and buoyancy field $(u_\nu',p_\nu,b_\nu)$ are solutions of the system

\begin{align}
 \partial_t u_\nu'&= -\partial_{x'}p_\nu -b_\nu \cos \phi +\nu \partial_{z'z'}u_\nu' \label{equationaprojetersurx}\\
0 &=-\partial_{z'}p_\nu +b_\nu \sin\phi   \label{equationaprojetersurz}\\
\partial_t b_\nu&= N^2 u_\nu'\cos\phi, \label{dtb'}
\end{align}
where we used $w_\nu = -u'_\nu \cos\phi$ (since $\vec{u}_\nu$ is parallel to $\vec{e}_{x'}$).
Equation~\eqref{equationaprojetersurz} leads to $p_\nu\sim d\,b_\nu$, thus the term $\partial_{x'}p_\nu$ in \eqref{equationaprojetersurx} can be neglected at zeroth order in $d/\lambda$.

As in the previous subsection, taking the ansatz  $a = \mathrm{Re} \left[ \underline{a}\ e^{-i\omega_0 t} \right]$, 
\eqref{dtb'} leads to $\underline{b}_\nu=i \omega_0 \underline{u}_\nu'\cos\phi/\sin^2\theta$. 
{Using \eqref{equationaprojetersurx}, we finally \ch{get}
 
 \begin{align}
 i\,\underline{u}_\nu'=(d/\mu)^2\partial_{z'z'}\underline{u}_\nu',
  \end{align}
 by  introducing the geometrical factor $\mu =({\cos^2\phi}/{\sin^2\theta}-1)^{1/2}$ and {$d$ the thickness of the boundary layer which is again equal to $\sqrt{\nu/\omega_0}$}. 
As this equation has to be solved with the boundary conditions $\lim_{z'\rightarrow-\infty}\underline{u}'_\nu=0$ and $\underline{u}'_\nu(x',z'=0)=-\underline{u}_{\mathrm{0}}(x') $,
where  ${u}_{\mathrm{0}}'(x')$ 
 is the inviscid along-wall velocity field, the solution reads

\begin{equation}
\underline{u}_\nu'(x',z')=
\begin{cases}
-\underline{u}_{\mathrm{0}}'(x') \exp\left({\frac{1+i}{\sqrt{2}}\mu_{{j}} z'/d}\right)  , &\mathrm{for }\  \phi=\alpha,0\ (j=1,3),\\
-\underline{u}_{\mathrm{0}}'(x') \exp\left({\frac{1-i}{\sqrt{2}} z'/d}\right) , & \mathrm{for }\  \phi=\pi/2\  (j=4).
\end{cases}
\end{equation}
After some calculations, one gets 

\begin{align}
\langle p^{[j]}_\perp\rangle_{{T_0}}
=  \frac{\nu \sqrt{2} \vert \mu_j\vert}{4{\cal A}d} \ \int\mathrm{d}x'  \vert \underline{u}_{\mathrm{0}}^'(x') \vert ^2 . \label{eqEpsPerp} 
\end{align}

\begin{figure}
\begin{center}
\includegraphics[scale=0.7]{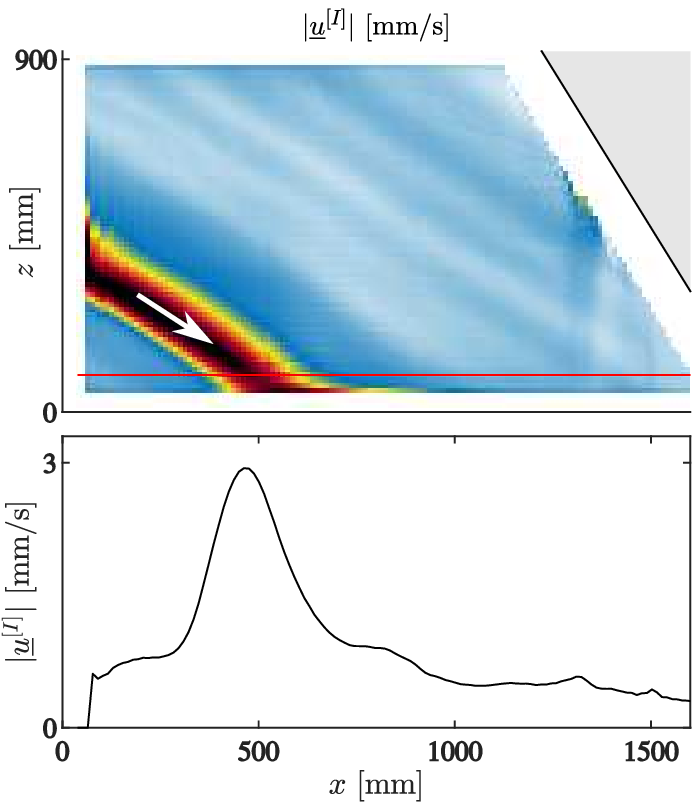} 
\includegraphics[scale=0.7]{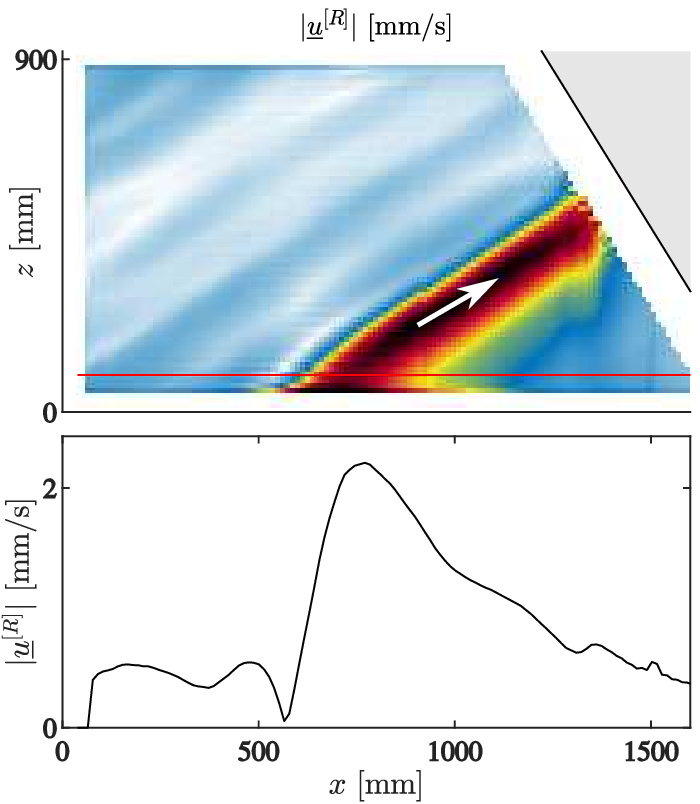} 
\caption{Incident {(left panels)} and reflected {(right panels)} velocity amplitudes {(top panels)} along the direction of the bottom boundary, obtained {at $t=50\ T_0$} thanks to Hilbert filtering. 
{The bottom panels correspond to the horizontal cut at the altitude $z=100$~mm shown in the top panels by the red horizontal lines.}
{Experimental parameters are those given in the caption of figure~\ref{figure_typ_PIV}}. 
}
\label{figurerebond}
\end{center}
\end{figure}
The amplitude of the inviscid along-wall velocity field $u_{\mathrm{0}}'(x')$ is required to proceed and compute the integral~\eqref{eqEpsPerp}. 
{As $u_{\nu}'$ vanishes outside from the boundary layer, }the PIV measurements give a good estimate of the inviscid velocity field.
 Furthermore, rather than measuring  $u_{\mathrm{0}}'(x')$
 directly on the PIV field by zooming on the region as close as possible to the wall, we use the relation $\vec{u} = \vec{u}^{[I]}+\vec{u}^{[R]}$, where $\vec{u}^{[I]}$ 
and $\vec{u}^{[R]}$ do not depend strongly on $z'$ along the rays which allows us to measure them where we want in the region next to the wall.
To get $\vec{u}^{[I]}$ and $\vec{u}^{[R]}$, we use the Hilbert transform~\citep{MGD2008} that allows one to distinguish the different branches of the attractor by restricting to 
internal waves propagating in a given direction. Because of experimental errors, $\vec{u}^{[I]}+\vec{u}^{[R]}$ is not perfectly parallel to the wall but {this is solved} by  computing its projection

\begin{align}
u_0' &= \left(\vec{u}^{[I]} +\vec{u}^{[R]} \right) \cdot \vec{e}_{x'} = \left( u^{[I]}  + u^{[R]}  \right) \sin{\phi} -  \left( w^{[I]}  + w^{[R]}  \right) \cos{\phi}.
 \label{eq_u_FS_XP}
 \end{align}
{For visualisation purpose}, $\vec{u}^{[I]}$ and $\vec{u}^{[R]}$ are {experimentally} evaluated at a small {but finite} distance {from} the boundary,  $z'=100$~mm. Because of this, one has to shift both functions  so that their maxima coincide as it is the case on the wall $z'=0$.
 Figure~\ref{figurerebond} shows an example of the incident and reflected beams, in the case of the bottom reflection. {From this profile,
 one can compute   
$  \langle p _\perp\rangle_{T_0}=  \langle p^{[1]} _\perp\rangle_{T_0}+  \langle p^{[3]}_\perp\rangle_{T_0}+\langle p^{[4]}_\perp\rangle_{T_0}$
using ~\eqref{eqEpsPerp}. 
As we will see in next section, this sum  {is of the same order of magnitude of the dissipation in the bulk} and is ten times smaller that the dissipation taking place in the parallel boundary layers. This result is expected since the volume of the parallel boundary layers is much larger due to the small width of the tank. Varying $W$ clearly changes the importance of the parallel boundary layers compared to the perpendicular ones{, as will be shown in~\S~\ref{3D_tank}}.
 
We will now apply these methods to measure the different terms in the energy budget for the  attractors in several regimes.

\section{Results\label{sec:results}}
\subsection{Linear regime}\label{Linearregime}

To validate the quantitative estimation of the different terms of the energy balance, a complete analysis is performed for an attractor in the linear regime. This is done with a small forcing amplitude ($a_0=2$~mm) using the following experimental scenario. The stratified fluid{ being initially} at rest, the wave maker is  turned on at $t=0$. After few forcing periods $T_0$,  the injected energy rate becomes stationary  {as anticipated in figure~\ref{figure_pression}(b)}. Finally, after 60~$T_0$, the generator is turned off to observe the decay of the attractor. The  power spectrum of the velocity field is displayed in figure~\ref{figure_tf_mes2}(a). We see that most of the energy is located near the forcing frequency $\omega_0/N=0.52$: the signal is {then} filtered around the forcing frequency to remove the external noise which could increase the errors in the estimation of the dissipative terms. 

Figure~\ref{figure_tf_mes2}(b) presents the time evolution of the different terms  {involved} in the energy balance:
 the injected energy $p_{\textrm{inj}}$ 
  and the three contributions of dissipation $p_{\textrm{bulk}}$, $p_{{\perp}}$ and~$p_{{\parallel}}$. 
 As mentioned in \S~\ref{sec:injected_power},  $p_{\textrm{inj}}$ is computed at $x_0\simeq5$~cm and therefore the dissipation occurring near the generator should
 not be taken into account. For this reason, we omit  $p_\perp^{[3]}$ and correct $p_{\textrm{bulk}}$ and~$p_{{\parallel}}$ accordingly.
 We  see that the variations of the total energy are very well given by the difference of the injected power with the three dissipative terms. After approximately 30 periods, a steady state has been reached. Without any adjustable parameters, the difference between the estimated dissipation and injected power is then less than {$10\%$}}, showing therefore a good agreement within experimental errors. 
As emphasized by figure~\ref{figure_tf_mes2}(b), this good agreement holds also in non-stationary situations, when one turns on or off the generator. 
Indeed, both curves have similar variations just after the switching time{s} $t=${0 and} $60~T_0$, due to the {changes in injected power. Notice that the fluctuations appearing on both curves just before $t=60~T_0$ are due to the applied frequency filter, thus limiting sharp changes in the signals.

Similar experiments  have been realized with larger amplitudes and therefore larger injected power. For each experiment, the signal is filtered around {the forcing frequency}~$\omega_0$ and the dissipation, measured in the steady-state, is plotted as a function of the injected power in figure~\ref{comp_pinj_pdiss} {(dotted line)}. 
The agreement is excellent at low amplitude but
the deviation between the dissipation at $\omega_0$ and the injected power increases when considering larger {amplitudes}.
This is expected since nonlinear effects can no longer be neglected: one has to consider other frequencies as discussed in the next-subsection.

\begin{figure}
\begin{center}
(a)\includegraphics[width=0.45\linewidth]{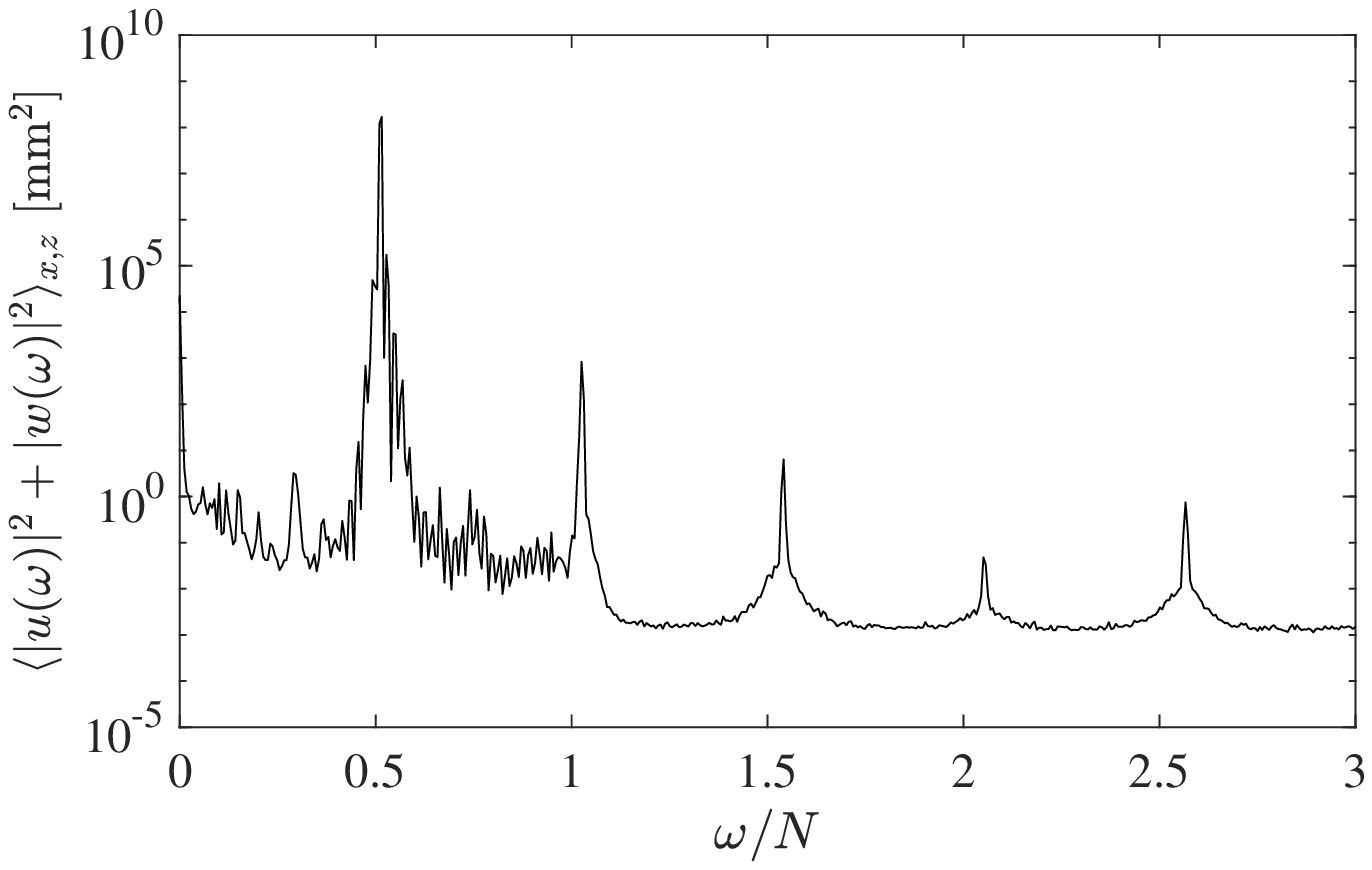}
(b)\includegraphics[width=0.45\linewidth]{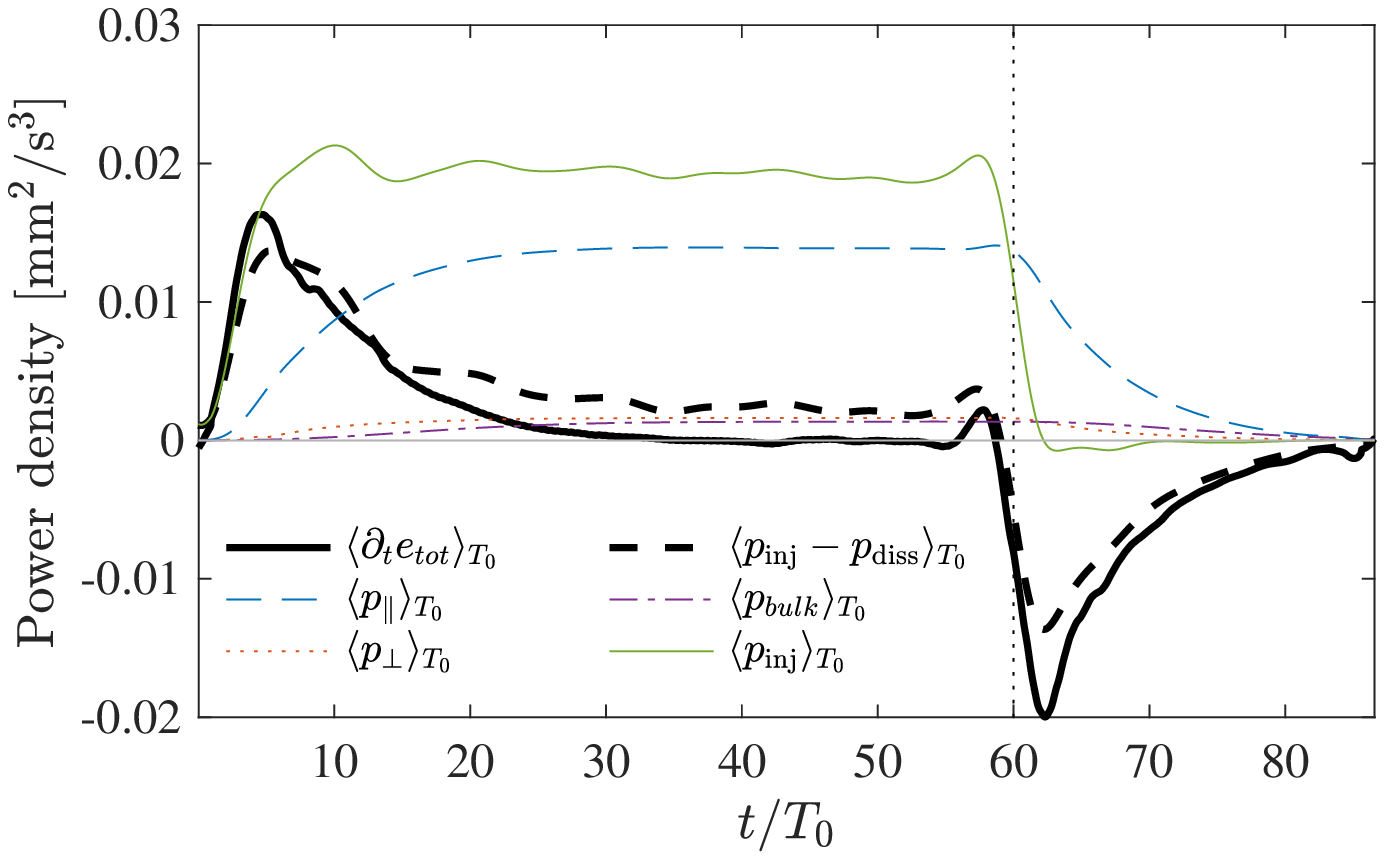}
\caption{Linear regime. For both figures, experimental parameters are those given in the caption of figure~\ref{figure_typ_PIV} with a forcing amplitude $a_0 =2$~mm. (a) Power spectrum as a function of the frequency normalized by the buoyancy frequency~$N$ (averaged over the area ${\cal A}$ and in logarithmic scale). Only waves with frequencies below this threshold value are \ch{propagating}, but one clearly sees several harmonics. (b) Evolution in time of the derivative of the total energy, the injected power and the different dissipative terms.
The {velocity} field has been filtered around the forcing frequency {$\omega_0$} before computing these quantities. {Oscillations at $\omega_0$ have then been removed using a moving average based on their extrema.} 
The generator has been switched on at $t=0$ while turned off at $t=60~T_0$, as emphasized by the vertical dotted line.}
\label{figure_tf_mes2}
\end{center}
\end{figure}

\begin{figure}
\begin{center}
\includegraphics[width=0.55\linewidth]{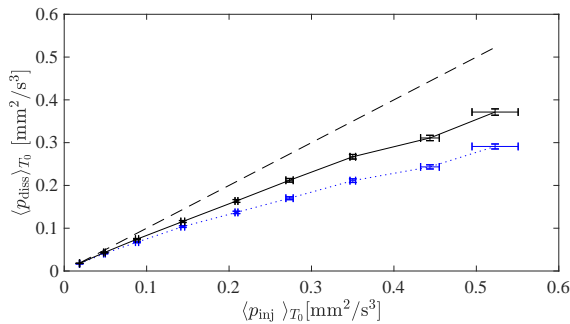} 
\caption{Full dissipation as a function of the measured injected power in the stationary regime. The blue dotted line corresponds to the velocity field filtered around the excitation frequency~$\omega_0$. 
The black line takes into account all frequencies
 below the buoyancy frequency~$N$. 
 \ch{The $x$-axis error bars are computed through the fluctuations of $\langle p_{\mathrm{inj}} \rangle_{T_0}$ in time, while  $y$-axis error bars come from the uncertainty of the PIV measurement of velocity field}.
The dashed \ch{line} corresponds to the {expected} relation $\langle p_{\mathrm{inj}}\rangle_{T_0} = \langle p_{\mathrm{diss}}\rangle_{T_0}$.
}
\label{comp_pinj_pdiss}
\end{center}
\end{figure}

\subsection{Weakly nonlinear regime}
\label{sec:weakly_non_linear}
By increasing the amplitude of the forcing, with values from $2$ to $10$ mm, the nonlinear regime \ch{is reached}.  {As underlined by~\cite{BrouzetEPL2016}, the setup could model at large amplitude a cascade of triadic interactions transferring energy to different frequencies and smaller scales. For all amplitudes, a stationary state, defined as the absence of time-variation of the total energy, is observed meaning that the injected power is balanced by the dissipation.}

We compute the different terms of \eqref{eq_bilan} as we did in the linear case. The injected power is computed exactly in the same way, with a prior filtering of the velocity field at the forcing frequency. Indeed, since the generator oscillates exactly at $\omega_0$, one has $p_{\mathrm{inj}}(\omega \neq \omega_0)=0$ as verified experimentally. \ch{Some non linear corrections to the estimation of the pressure gradient have been computed and appeared to be very small (less than 1\%). They are therefore neglected in the following.}
  To evaluate the parallel dissipation $p_{\parallel}$, a similar method than the one described in \S~\ref{sec:diss_parallel} is used. To do so, we assume that the linear approximation used to obtain the velocity field in the boundary layer is still valid. Equations~\eqref{eq:theory_velocity_long1} and~\eqref{eq:theory_velocity_long2} are therefore multiplied by the time Fourier transform of the velocity field, low-pass filtered with a cut-off frequency equal to the buoyancy frequency~$N$. The real part of the inverse time Fourier transform of the obtained signal is finally used to get the value of $\partial_y \vec{u}(x,z,t)$ and $p_{\parallel}$ in the boundary layers for all waves with frequency smaller than~$N$ present in the domain.

As we showed that $p_\perp \simeq 0.1 \ p_\parallel$, we neglect the perpendicular dissipation due to the secondary waves, and we will consider $p_\perp\simeq p_\perp(\omega_0)$. On the contrary, as secondary waves {usually} present smaller length scales, one cannot \textit{a priori} neglect their contribution to the bulk dissipation, which is measured here by computing~\eqref{eq_epsilon_bulk}  with only a band-pass filter to remove noise above the buoyancy frequency.

The black line of figure~\ref{comp_pinj_pdiss} shows,  in the stationary regime, the dissipation  measured in this way 
as a function of the  injected power. The agreement with the expected relation $\langle p_{\mathrm{inj}}\rangle = \langle p_{\mathrm{diss}}\rangle_{T_0}$ represented by a dashed line has been significantly improved with respect to the one obtained when considering only  the dissipation  at~$\omega_0$.  {For very large amplitudes, the dissipation associated with the forcing $\omega_0$ represents only half of the injected power. Moreover,  we have quantitatively shown that the secondary waves forced by successive triadic interactions represent around 16\%  of the {injected power} in the strongest nonlinear regime.

 {To go into more details, we will describe the time evolution of the different terms of the energy balance in the nonlinear regime.} A typical power spectrum, for an intermediate forcing amplitude $a_0=5$ mm and an injected power of $1.9\times 10^{-2}$~mm$^2$/s$^3$, is shown in figure~\ref{figure_tf_Ec_mes5}(a). {While} the wave-field is strongly dominated by the forcing frequency $\omega_0$, one clearly sees several other peaks. 
With this tool, one can investigate the exchange of energy between the different components of the velocity field. {F}igure~\ref{figure_tf_Ec_mes5}(b) shows the evolution of the energies of the five most important components identified in  figure~\ref{figure_tf_Ec_mes5}(a).
After the generator has been turned on at $t=0$, the energy of the primary wave~$\omega_0$ increases. 
Then, approximatively $25$ periods later (that corresponds to the typical time scale for the appearance of the TRI),  some energy loss of the primary wave is observed while the energy of the secondary waves increases indicating an energy transfer. 
\ch{Some of them present some decrease of energy, which seems to show a complex interaction between the secondary-waves.}
 As can be seen in figure~\ref{comp_pinj_pdiss}, the dissipation due to the primary wave slightly underestimates the injected power. For this weakly nonlinear case, the energy of the secondary waves is very small, and so is their dissipation. Even if it is of the same order of magnitudes than error bars, it seems that their presence explain the difference between the injected power and the dissipation due to the forcing frequency.

\begin{figure}
\begin{center}
\includegraphics[scale=0.4]{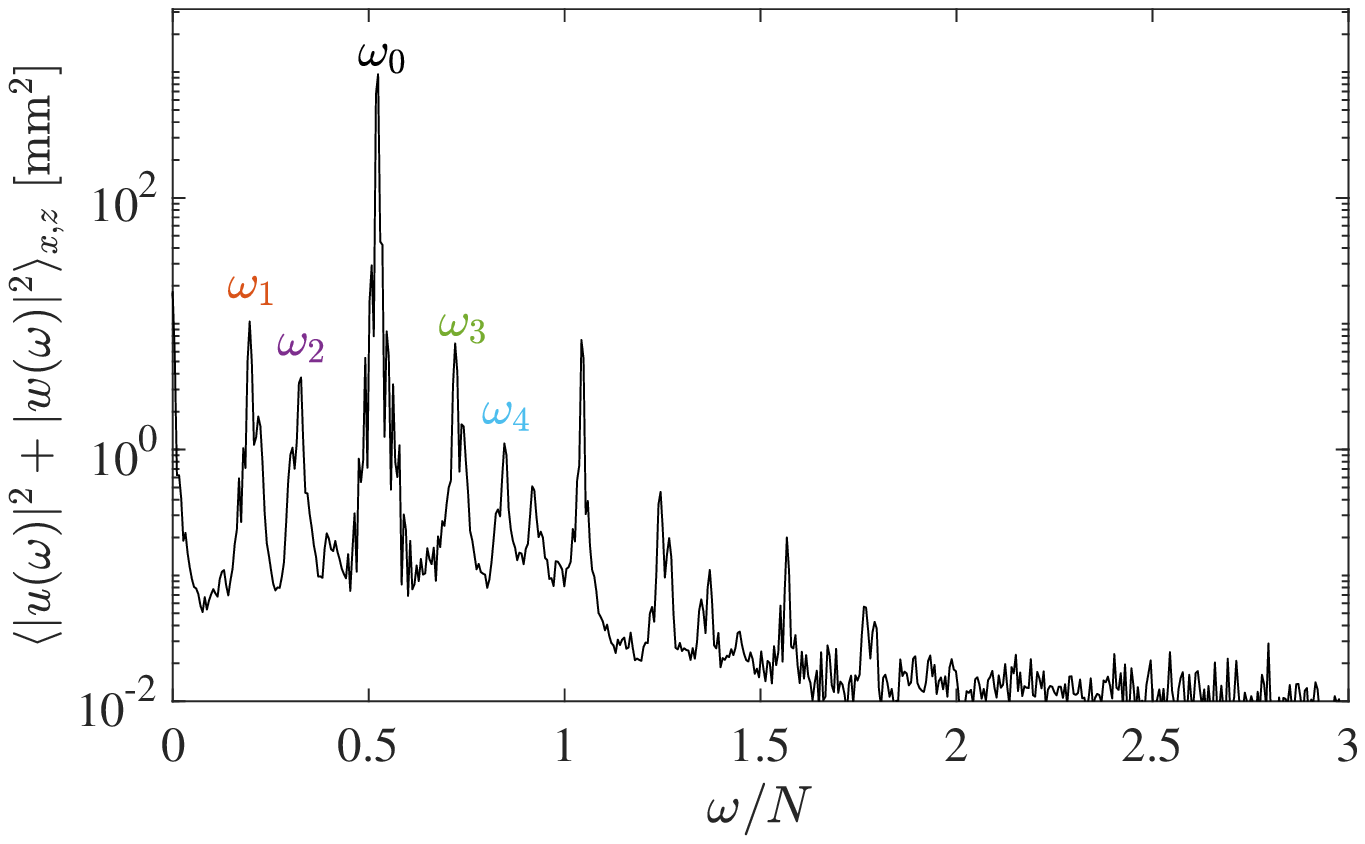} 
\includegraphics[scale=0.4]{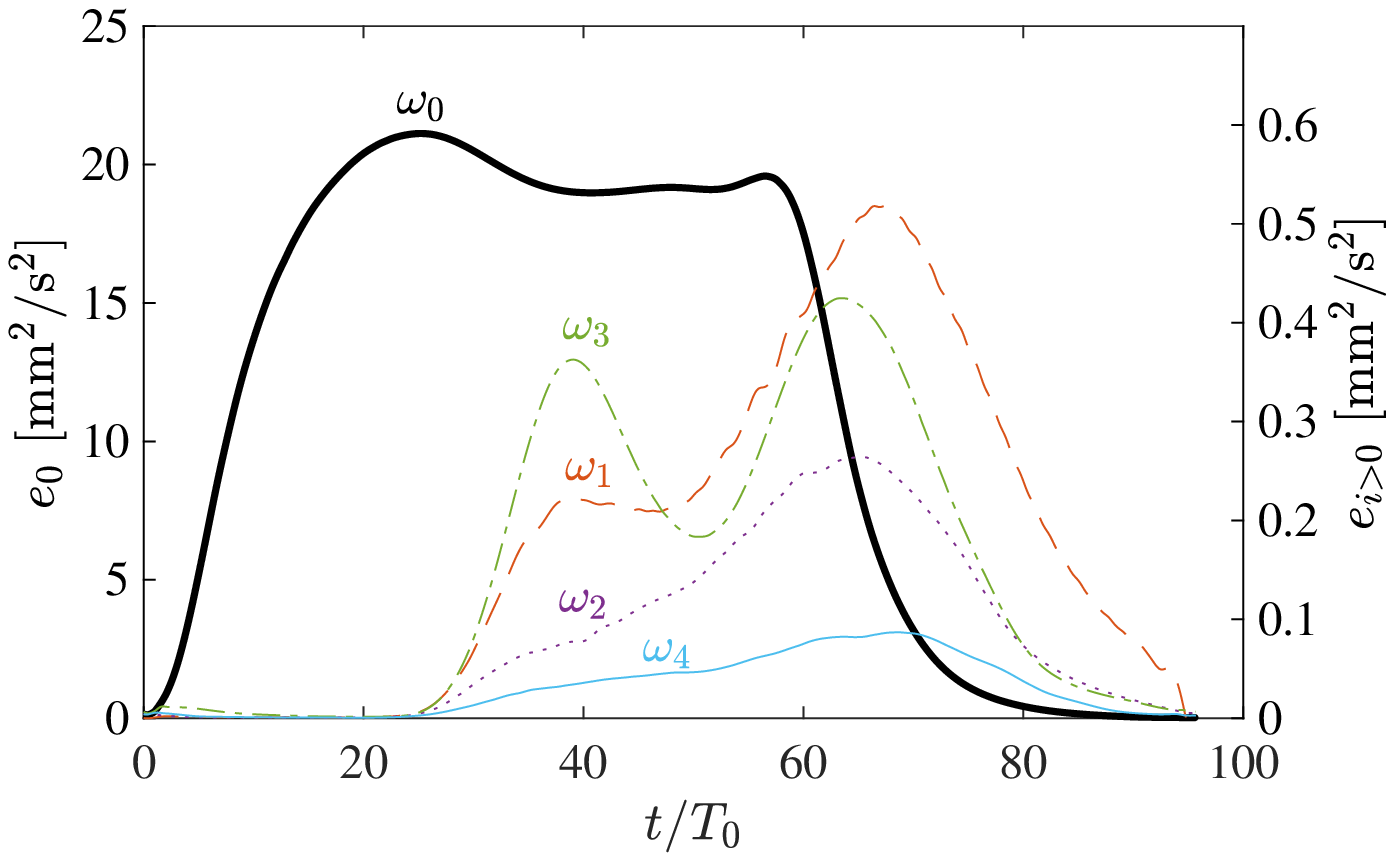} 
\caption{Weakly nonlinear regime. (a) 
Power spectrum as a function of the frequency normalized by the buoyancy frequency~$N$ (averaged over the area ${\cal A}$ and in logarithmic scale).
Experimental parameters are those given in the caption of figure~\ref{figure_typ_PIV} except $a_0=5$ mm. The five main frequencies have been labelled {$\omega_i$ with $i=0$ to 4}.
(b) Evolution of the energy of the five main components of the velocity field. 
Note the difference in the energetic scale between $e_0$ the energy of the forcing frequency  $\omega_0$ on the left {$y$-axis}
and, on the right, the other ones $e_i$ with $i>0$ for each $\omega_i$.
 The generator is switched on at $t=0$ while turned off at $t=60~T_0$.}
\label{figure_tf_Ec_mes5}
\end{center}
\end{figure}

 To confirm this hypothesis, we look into more details to the case with the largest amplitude $a=10$~mm, which correspond{s} to a stronger nonlinear regime with a rich multi-peak spectrum. Figure~\ref{figure_tf_Ec_mes10} displays the time evolution of the different terms {involved} in the energy balance in which we have separated the dissipative part at the forcing frequency and the part due to the secondary waves. The agreement is very good at the beginning of the experiment just after switching on the generator. At about $20~T_0$, a stationary state is reached where the time derivative of the total energy $\partial_t e_{\textrm{tot}}$ vanishes with some oscillations. At $15~T_0$, the dissipation due to the forcing frequency decreases and is substituted by dissipation due to secondary waves. The latter finally reaches 
 $16\%$ of the injected power.  
 It is important to note that the bulk dissipation {for secondary waves constitutes}$30$\% {of total} secondary{-}waves {dissipation} and {is} therefore much more important than for the primary wave, in which $p_{\textrm{bulk}}(\omega_0)\approx 10\%p_{\textrm{diss}}(\omega_0)$. This result is expected since the typical length scales for the secondary waves are smaller. Even though the estimation of the dissipation {due to} the forcing frequency and the secondary waves gives an interesting estimation of the injected energy within the system, figures~\ref{comp_pinj_pdiss} and~\ref{figure_tf_Ec_mes10} show that the part of the dissipation not taken into account represents a quarter of the total dissipation. It is more important when nonlinearities are stronger. Different hypothesis may be given to explain this missing part in the dissipation. (i) In our model, we did not take into account frequencies larger than~$N$ {for all dissipation} {n}or polychromaticity in the boundary layers when we estimate{d} the perpendicular  dissipation. The reflection at the free surface may also be the origin of some dissipation, since the boundary conditions is in reality between no slip and free slip \ch{due to surface contamination~{\citep{campagne2018}}}. 
(ii)  The average over the forcing period~$T_0$ of some terms in \eqref{equationpbypartsvolu} is not vanishing  anymore in such a polychromatic case. 
 (iii) The bulk dissipation may be underestimated. Indeed, the \ch{smallest} scale we have access to is given by the resolution of the PIV measurements. Due to nonlinearities, some energy may be transferred to scales smaller than $10$~mm and is not measured. (iv) Motions or dependence in the $y$-direction are not taken into account. 
 Secondary waves are assumed here to keep the two-dimensional geometry,  which is not certain. \ch{(v) We see in figure~\ref{figure_tf_mes2} that some energy goes into a mean flow generated in the boundary layers~\citep{Horne2019}.
 While the associated dissipation in the bulk is taken into account, its counterpart in the boundary layers is not.} (vi) Finally, part of the energy may be taken to increase the potential energy of the fluid, which is related to a homogeneisation of the density through mixing.

\begin{figure}
\begin{center}
\includegraphics[scale=0.4]{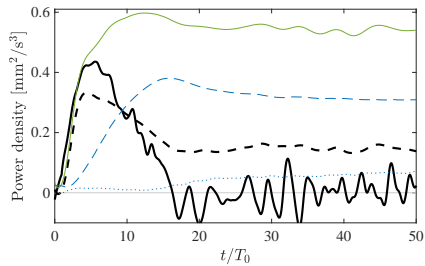} 
\caption{Nonlinear regime. Evolution in time of the derivative of the total energy $\partial_t e_{\textrm{tot}}$ (thick black line), the injected power $p_{\textrm{inj}}$ (green solid line), the dissipation associated with the forcing frequency $p_{\textrm{diss}}(\omega_0)$ (thin blue dashed line), the dissipation associated with other frequencies smaller than $N$, $p_{\textrm{diss}}(\omega<N,\omega\neq\omega_0)$ (blue dotted line) and finally the difference $p_{\textrm{inj}}-p_{\textrm{diss}}$ (thick black dashed line). Experimental parameters are those given in the caption of figure~\ref{figure_typ_PIV} except $a_0=10$ mm.}
\label{figure_tf_Ec_mes10}
\end{center}
\end{figure}

\subsection{Three dimensional tank}
\label{3D_tank}

As already underlined, the geometry of the domain has an  effect on the importance of the different dissipative terms. Indeed, neglecting the feedback of the dissipation on the attractor, the bulk part as well as the perpendicular part depend linearly on the {thickness} of the tank, while the parallel part is independent {of} $W$.
To investigate a regime where dissipation {is} less dominated by {the dissipation due to lateral walls, i.e. with a relatively lower} $p_\parallel$,
we ran an experiment {in a linear regime} in a different tank of size  {$L\times W\times H=$} $570 \times 800 \times 320 \ \mathrm{mm^3}$ with a comparable slope of $\alpha=29^\circ$; \ch{the wave maker is 150~mm wide}. As shown in~\cite{Grimaud2018} in which the \ch{details} of the experimental set-up is given,
the hypothesis of the two-dimensionality of the flow still holds in {this tank within} the linear and stationary regime. Note that the perpendicular dissipation  $p_\perp^{[3]}$ on the left has now to be taken into account since the wave maker does not occupy the total width of the tank. As shown in figure~\ref{bilan_insta_3D}, 
we found that in this case the three dissipation rates are more comparable in \ch{magnitude}, since $W$ is now $5$ times larger {while $L$ and $H$ are both about 3 times smaller}. 
 The agreement between the measured quantities and the theory is again very good both in the non stationary and the stationary regimes.

\begin{figure}
\begin{center}
\includegraphics[scale=0.4]{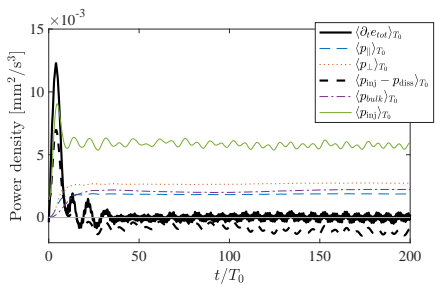}
\caption{
Evolution of the derivative of total energy, injected power and the different dissipative terms in time, for a linear experiment with forcing amplitude $a_0=5$ mm in the 3D tank. \ch{Note the $10^{-3}$ factor on the $y$-axis.} The field had been filtered around the forcing frequency before computing these quantities.}
\label{bilan_insta_3D}
\end{center}
\end{figure}

\section{Conclusions\label{sec:conclusion}}
The goal of the work presented in this paper was to investigate the energy budget of a two-dimensional internal wave attractor using experiments. In the linear regime, we have shown that 
it is possible to  simultaneously measure the energy flux injected into the system and the dissipation rate using 2D-2C PIV measurements. The pressure field, usually not accessible in laboratory or field measurements, is computed 
from the velocity field and gives access to the injected power $p_{\textrm{inj}}$. The dissipation is disentangled into three different terms following the work of~\cite{Beckebanze2018}: the bulk part is directly calculated 
from the velocity field, while the parallel and perpendicular parts due to the dissipation in the boundary layers, in the lateral walls and at the reflection of the internal wave beam, are quantified 
using a linear \ch{modeling}  of  the boundary layer. We have shown that this procedure \ch{captures} without any adjustable parameters the complete energy balance in the stationary state as well as in the transient regime. 

{M}oreover, two different experimental setups have been used to quantify the importance of the parallel dissipation compared to the bulk dissipation. 
The ratio depends on the distance between the lateral walls and on the thickness of the boundary layers compared with the typical wavelength of the internal waves within the attractor. Using similar scaling arguments, the ratio between the bulk and the perpendicular dissipations depends on the thickness of the boundary layers and the typical size of the regions where a reflection occurs, which is close to the perimeter of the domain compared with the typical wavelength of the internal waves within the attractor.

Finally, we have shown that our analysis developed for the linear case is still valid in the nonlinear regime. Even if we cannot completely capture the total dissipation in the tank, we have quantified that the secondary waves generated through triadic resonant interactions have clearly a non negligible part in the dissipation rate.
As the difference between dissipation and variation of energy of a given component of the wave field gives access to its energy input, the analysis presented here could be  part of the  investigation of  nonlinear energy transfers between  different waves.
 \ch{It is also important to emphasize that the method tested here on internal wave attractors can be generalized straightforwardly to any quasi two-dimensional stratified flow.}

\begin{acknowledgments}
{\bf {Acknowledgments}}
We thank E. Ermanyuk, B. Gallet,  and I. Sibgatullin for insightful discussions. 
 This work was supported by the LABEX iMUST (ANR-10-LABX-0064) of Université de Lyon, within the program “Investissements d’Avenir” (ANR-11-IDEX-0007), operated by the French National Research Agency (ANR), France. This work has been supported by the ANR, France through grant ANR-17-CE30-0003 (DisET). This work has been achieved thanks to the resources of PSMN from ENS de Lyon.
\end{acknowledgments}

\end{document}